\documentclass[12pt]{article}

\usepackage{amsmath,amssymb}
\usepackage{graphicx}

\makeatletter
 
  \@addtoreset{equation}{section}
 \makeatother

\newcommand{\be}{\begin{equation}}
\newcommand{\ee}{\end{equation}}
\newcommand{\bea}{\begin{eqnarray}}
\newcommand{\eea}{\end{eqnarray}}
\newcommand{\beann}{\begin{eqnarray*}}
\newcommand{\eeann}{\end{eqnarray*}}
\newcommand{\nn}{\nonumber}
\newcommand{\ba}{\begin{array}}
\newcommand{\ea}{\end{array}}

\newcommand{\Tr}{{\rm Tr}\,}

\newcommand{\R}{\mathbb{R}}

\newcommand{\kv}{{\bf k}}
\newcommand{\av}{{\bf a}}
\newcommand{\e}{\epsilon}
\newcommand{\CP}{\mathbb{C}{\rm P}}

\newcommand{\N}{{\cal N}}
\newcommand{\pint}{-\hspace{-13pt}\int}

\renewcommand{\th}{\widetilde{h}}

\begin{document}

%%
%% TITLE
%%
\begin{titlepage}

%% Set the number of the title with 0
\setcounter{page}{0}
\renewcommand{\thefootnote}{\fnsymbol{footnote}}

\begin{flushright}
hep-th/0406191\\
RIKEN-TH-27
\end{flushright}

\vspace{20mm}
\begin{center}
{\large\bf
Large $N$ limit of 2D Yang-Mills Theory and Instanton Counting}

\vspace{20mm}
{
Toshihiro Matsuo\footnote{{\tt tmatsuo@riken.jp}},
So Matsuura\footnote{{\tt matsuso@riken.jp}} and
Kazutoshi Ohta\footnote{{\tt k-ohta@riken.jp}}}\\
\vspace{10mm}

{\em Theoretical Physics Laboratory\\
The Institute of Physical and Chemical Research (RIKEN)\\
2-1 Hirosawa, Wako\\
Saitama 351-0198, JAPAN}\\

\end{center}

\vspace{20mm}
\centerline{{\bf Abstract}}
\vspace{10mm}

We examine the two-dimensional $U(N)$ Yang-Mills theory 
by using the technique of random partitions. 
We show that the large $N$ limit of the partition function 
of the 2D Yang-Mills theory on $S^2$ 
reproduces the instanton counting of 4D $\N=2$ 
supersymmetric gauge theories introduced by Nekrasov.
We also discuss that we can take the ``double scaling limit'' by fixing 
the product of the $N$ and cell size in Young diagrams, 
and the effective action given by Douglas and Kazakov 
is naturally obtained by taking this limit. 
We give an interpretation for our result from the view point 
of the superstring theory by considering a brane configuration 
that realizes 4D $\N=2$ supersymmetric gauge theories.

\end{titlepage}
\newpage

\renewcommand{\thefootnote}{\arabic{footnote}}
\setcounter{footnote}{0}

\section{Introduction}

It has been recently recognized that ``duality'' 
is a quite important idea to understand 
non-perturbative aspects of gauge theories.  
For example, 
the duality between gauge theories and 
string theories \cite{'tHooft:1974jz} is realized as 
the AdS/CFT correspondence 
\cite{Maldacena:1998re,Witten:1998qj,Gubser:1998bc} 
(For review, see \cite{Aharony:1999ti,Fukuma:2002sb}), 
the large $N$ reduction of gauge theories realizes 
the gauge/matrix correspondence 
\cite{Eguchi:1982nm,Gross:1982at,Bhanot:1982sh,Das:1982ux}
and large $N$ matrix models have been proposed as candidates 
for non-perturbative definition of the string/M theory
\cite{Banks:1997vh,Ishibashi:1997xs}, 
and recently, it was found that 
$\N=1$ supersymmetric gauge theories closely related to 
$c=0$ matrix models \cite{Dijkgraaf:2002dh}. 

Among studies of duality in lower dimensional theories, 
two-dimensional (2D) Yang-Mills theory has played important 
roles to examine the nature of the gauge/string correspondence. 
This theory is exactly solvable \cite{Migdal:1975zg} and 
the partition function of 2D $U(N)$ ($SU(N)$) gauge theory
on an arbitrary orientable manifold of genus $G$ with area $A$ 
is exactly given by a sum over all irreducible representations 
\cite{Rusakov:1990rs};
\be
Z_G=\sum_{R} (\dim R)^{2-2G}e^{-\frac{\lambda A}{2N} C_2(R)},
\label{partition function}
\ee
where $\dim R$ and $C_2(R)$ are a dimension and quadratic Casimir 
of the representation $R$. 
Gross and Taylor have attempted in their seminal works 
\cite{Gross:1993tu,Gross:1993hu,Gross:1993yt}
to uncover the relationship between 
the 2D Yang-Mills theory and a string theory. 
(For review, see Ref.\,\cite{Cordes:1995fc}.)
Not only the gauge/string correspondence, 
the equivalence with a $c=0$ matrix model \cite{Douglas:1993ii} 
and a $c=1$ matrix model \cite{Minahan:1993np} 
has been studied. 
As recent developments, a correspondence between the finite $N$ 
2D Yang-Mills and a string theory has been discussed 
\cite{Lelli:2002gr,Matsuo:2004nn}%
\footnote{For early work for this subject, see 
Ref.\,\cite{Baez:1994gk}.},  
and it has been pointed out that 
2D Yang-Mills theory 
relates to 4D black holes \cite{Vafa:2004qa} 
and random walks \cite{deHaro:2004id}. 

In this article, 
we analyze 2D $U(N)$ Yang-Mills theory 
from the point of view of random partitions 
and discuss the relation to 
the instanton counting of 4D 
$\N=2$ supersymmetric gauge theories 
discovered by Nekrasov \cite{Nekrasov:2003rj}.
The authors of Ref.\,\cite{Nekrasov:2003rj} have used 
the technique of a summing over random partitions. 
This technique is powerful enough 
when partitions or Young diagrams are concerned 
to carry out the exact calculation of the instanton counting. 
However, in spite of the beautiful structure of 
the obtained instanton partition function, 
the physical meaning of the relationship between 
the instanton counting and the random partitions 
is not clear yet.

We show that we can rewrite the partition function 
(\ref{partition function}) in the language of random partitions. 
In particular, we argue that the large $N$ limit 
of the partition function (\ref{partition function}) 
reproduces the instanton counting of a 4D 
${\cal N}=2$ supersymmetric gauge theory given in 
Ref.\,\cite{Nekrasov:2003af} 
by a deformation corresponding to the breaking of the gauge symmetry, 
$U(N)\to U(N_1)\times\cdots\times U(N_r)$. 
We also argue that we can take a ``double scaling limit''
by fixing the product of $N$ and the size of boxed of 
Young diagrams, 
which includes the naive large $N$ limit 
as a limit of the fixed parameter. 
We will see that this limit naturally reproduces the discussion 
by Douglas and Kazakov \cite{Douglas:1993ii}.

This paper is organized as follows. 
In the next section, we review the definition and some properties 
of the profile function that is convenient to express partitions. 
We show that the dimension and the quadratic Casimir of 
$U(N)$ can be rewritten using the profile function. 
In the section 3, we take the large $N$ limit of 
the 2D Yang-Mills theory and show that 
its partition function gives 
the instanton counting of a 4D 
${\cal N}=2$ supersymmetric gauge theory. 
In the section 4, we argue the ``double scaling limit'' 
of the partition function. 
In the section 5, we consider a brane configuration which 
realizes 4D ${\cal N}=2$ supersymmetric gauge theory 
and 2D Yang-Mills theory, 
and discuss a string theoretical interpretation for the result that 
2D Yang-Mills theory reproduces the instanton counting 
of the 4D theory. 
The section 6 is devoted to conclusions and discussions. 

\section{The profile function of Young diagram}

As mentioned in the introduction, the partition function of 
2D Yang-Mills theory is given by a sum over 
all irreducible representations of the gauge group, 
each of which corresponds to a Young diagram.  
In this section, we first review 
the definition and some properties of the ``profile function'' 
which expresses the Young diagram systematically, 
and rewrite the partition function (\ref{partition function}) 
using them. 

\begin{figure}[ht]
\begin{center}
\includegraphics[scale=0.8]{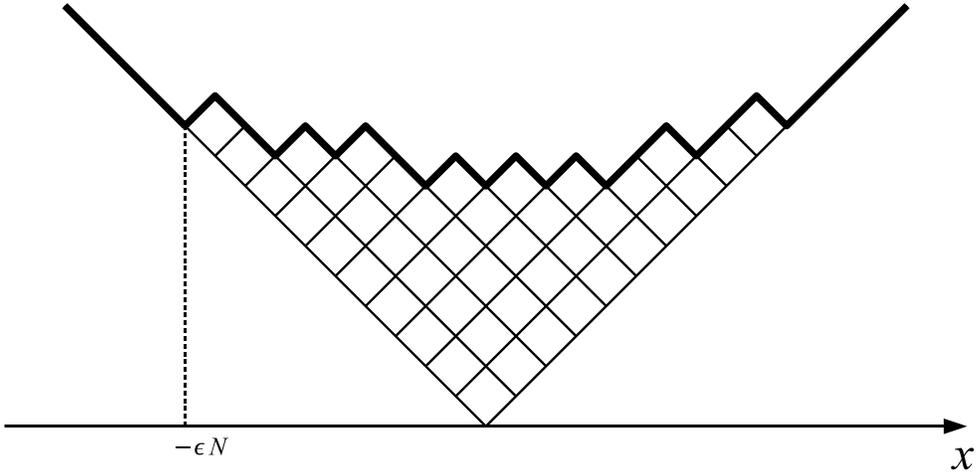}
\end{center}
\caption{The bold line on the rotated Young diagram represents
 the profile function. For $U(N)$ group, the Young diagram has a
 ``cut-off'' at $x=-\e N$ since the number
of row is restricted on less than $N$.}
\label{profile fig}
\end{figure}

Recalling the one-to-one correspondence between 
a Young diagram with $k$ boxes and a partition of $k$, 
$\left\{k_1,\cdots,k_L\right\}$ 
$(k_1\ge\cdots\ge k_L > 0, \,\, k_1+\cdots+k_L = k)$, 
we can identify a representation $R$ of $U(N)$ $(SU(N))$ group 
with the corresponding partition.
For the following discussion, 
we extend the definition of the partition as 
\begin{equation}
 \kv \equiv \left\{k_i\right\}_{i=1}^\infty, 
\end{equation}
with the restrictions, 
\begin{gather}
 \sum_{i=1}^{\infty}k_i = k, \nn\\
 k_1\ge\cdots\ge k_L > 0, \qquad
 k_j = 0 \,\, {\rm for} \,\, j>L.  
\end{gather}
Note that, for the irreducible representation of $U(N)$,
$L$ is less than or equal to $N$  
since $k_i$ expresses the length of $i$'s row of the diagram. 
Corresponding to the extended partition $\kv$, 
we define the profile function as; 
\cite{Nekrasov:2002qd}
\be
f_{\kv}(x|\e)=|x|
+\sum_{i=1}^{\infty}\left[
|x-\e(k_i-i+1)|
-|x-\e(k_i-i)|
+|x+\e i|
-|x+\e(i-1)|
\right]. 
\label{profile function}
\ee
As shown in Fig.\,\ref{profile fig}, 
in which we explicitly draw the profile function for the partition 
$\kv=\left[10,8,6,5,4,3,3,2,1,1\right]$, 
the standard shape of the Young diagram is rotated by 45 degree 
(and reflected) in the profile function. 

Let us also introduce a function $\gamma_{\e}(x)$ 
which obeys the difference equation, 
\be
\gamma_\e(x+\e)+\gamma_\e(x-\e)-2\gamma_\e(x)=\log(x). 
\ee
For later discussion, we write down some properties of 
the function $\gamma_\epsilon(x)$. 
(For detail, see the appendix of Ref.\,\cite{Nekrasov:2002qd}.) 
$\gamma_\epsilon(x)$ can be written as 
\begin{equation}
 \gamma_\epsilon(x) = \frac{d}{ds}\Bigg|_{s=0}
  \frac{1}{\Gamma(s)}\int_0^\infty \frac{dt}{t}t^s 
  \frac{e^{-tx}}
       {\left(e^{\epsilon t}-1\right)\left(e^{\epsilon t}+1\right)}, 
\end{equation}
up to linear functions of $x$. 
From this expression, we can read off the asymptotic expansion 
of $\gamma_\epsilon(x)$ for $\epsilon \to 0$; 
\begin{equation}
 \gamma_\epsilon(x) \equiv \sum_{g=0}^\infty 
  \epsilon^{2g-2}\gamma_g(x), 
\end{equation}
with
\begin{align}
 \gamma_0(x) &= \frac{1}{2}x^2 \log x - \frac{3}{4}x^2, \nn \\
 \gamma_1(x) &= -\frac{1}{12} \log x, \nn \\
 \gamma_2(x) &= -\frac{1}{240} \frac{1}{x^2},
\label{expansion of gamma} \\ 
 &\vdots \nn\\
 \gamma_g(x) &= \frac{B_{2g}}{2g(2g-2)}\frac{1}{x^{2g-2}}, \nn
\end{align}
where $B_{2g}$'s are the Bernoulli numbers.  

Combining $f_{\kv}(x|\e)$ and $\gamma_\e(x)$, 
we can prove the integration, 
\begin{align}
\exp\left\{
-\frac{1}{8}\pint dx dy\  f''_\kv(x|\e)f''_\kv(y|\e)\gamma_\e(x-y)
\right\}
=\prod_{1\leq i < j < \infty}
\frac{k_i-k_j+j-i}{j-i} 
=\prod_{\{\square\}}\frac{1}{h_\square} 
= \frac{d_\kv}{k!}, 
\label{infinite product}
\end{align}
where $\{\square\}$, $h_{\square}$ and $d_{\kv}$ 
denote the set of boxes of the Young diagram, 
the hook length about the box $\square$, 
and the dimension of the irreducible representation of 
the symmetry group $S_k$ corresponding to the partition $\kv$, 
respectively. 
From the first line to the second line, 
we have used the explicit expression of the second derivative 
of the profile function, 
\be
f''_{\kv}(x|\e)=2\delta(x)
+2\sum_{i=1}^{\infty}\left[
\delta(x-\e(k_i-i+1))
-\delta(x-\e(k_i-i))
+\delta(x+\e i)
-\delta(x+\e(i-1))
\right].
\ee
Notice that the product over $i,j$ runs infinitely and this is regarded
as the dimension of a representation $R$ associating with the given Young
diagram at the large $N$. This infinite product is well-defined measure on
the random partitions known as the Plancherel measure and will play a central
role in the following discussions.

From the above properties of the profile function, 
we can express the dimension 
of the representation $R$ as 
\begin{align}
\dim R &= 
\prod_{1 \leq i < j \leq N} \frac{k_i-k_j+j-i}{j-i}  \nn \\
&=\prod_{1 \leq i < j < \infty} \frac{k_i-k_j+j-i}{j-i} 
\prod_{{1 \leq i \leq N} \atop {N < j < \infty}}
\frac{j-i}{k_i+j-i} \nn\\
&=
\exp\left\{
-\frac{1}{8}\pint dx dy\  f''_\kv(x|\e)f''_\kv(y|\e)\gamma_\e(x-y)
+\frac{1}{2}\int dx\ f''_\kv(x|\e)
\left(\gamma_\e(x+\e N)-\gamma_\e(\e N)\right)\right\}, 
\end{align}
and the quadratic Casimir of the representation as 
\begin{align}
C_2(R)&=\sum_i\left(N k_i + k_i(k_i - 2i +1)\right)  \nn \\
&=\int dx f''_\kv(x|\e) \left(
\frac{N}{4\e^2}x^2
+\frac{1}{6\e^3}x^3
\right).
\label{Casimir potential}
\end{align}
Substituting them into (\ref{partition function}), 
we can rewrite the partition function 
of the 2D Yang-Mills theory 
 (\ref{partition function}) in terms of integrals on the profile function
\begin{align}
\label{profile partition}
Z_G = \sum_{k=1}^{\infty}\sum_{\kv \in Y_k^N}
\exp\biggl\{
&-\frac{2-2G}{8}\pint dx dy\  f''_\kv(x|\e)f''_\kv(y|\e)\gamma_\e(x-y) \nn\\
&+\frac{2-2G}{2}\int dx\ 
f''_\kv(x|\e)
\left(
\gamma_\e(x+\e N)-\gamma_\e(\e N)
\right)\\
&-\frac{\lambda A}{4 N}\int dx\ 
f''_\kv(x|\e)
\left(
\frac{N}{2\e^2} x^2
+\frac{1}{3\e^3} x^3
\right)
\biggr\}, \nn
\end{align}
where $Y_k^N$ stands for a set of the Young diagram of $U(N)$ representation
consisting with $k$ boxes.

This formulation of the partition function has a remnant of an usual
$c=0$ hermitian matrix model. 
Indeed, let us now consider 
the following matrix model with ``matter'' matrix as 
\be
Z_{{\rm mm}}=\int [d\Phi][dQ][d\tilde{Q}]
e^{-\frac{1}{g_s}\left(
V(\Phi) + \tilde{Q}(\Phi+m)Q
\right)},
\ee
where $\Phi$, $Q$ and $\tilde{Q}$ are $N\times N$, $N\times N_f$ and
 $N_f \times N$ hermitian matrices, respectively,
and $m$ represents mass of the matter.
Integrating $Q$ and $\tilde{Q}$ first and diagonalizing the matrix $\Phi$
into eigenvalues $\lambda_i$,
we find
\be
Z_{\rm mm} = \int \prod_i d\lambda_i
\frac{\prod_{i<j}(\lambda_i-\lambda_j)^2}{\prod_i(\lambda_i+m)^{N_f}}
e^{-\frac{1}{g_s}\sum_i V(\lambda_i)}.
\ee
If we define the eigenvalue density,
\be
\rho(x)=\frac{1}{N}\sum_i \delta(x-\lambda_i),
\ee
the effective action of the above matrix model becomes
\bea
S_{\rm eff}&=&-2 g_s N^2 \pint dx dy \ 
\rho(x)\rho(y)\log(x-y)\nn\\
&& \quad+ g_s N N_f \int dx \  \rho(x)\log(x+m)
+ N \int dx\,\rho(x) V(x).
\label{mm eff action}
\eea

2D Yang-Mills theory is a kind of discretized matrix
model as pointed out in Refs.\,\cite{Gross:1993tu, Douglas:1993ii}, 
and thus 
we expect that the matrix model technique is useful to solve
2D Yang-Mills theory. 
In fact, 
comparing the matrix model effective action (\ref{mm eff action})
with (\ref{profile partition}), we see the eigenvalue density
and logarithmic function corresponds to $f''_\kv(x|\e)$ and $\gamma_\e(x)$
naively. In addition, the Vandermonde determinant (the measure of the matrix
model) and potential relates to the (regularized) Plancherel measure and
Casimirs. We make a restriction on the number of rows of the Young diagram
in order to treat the finite $N$ case. This restriction causes introducing
the matter with a mass of $-\e N$ if we compare with the effect of the matter
with mass $m$ to the matrix model. This means that the effect of the finite $N$
behaves as a regularization at the mass scale of $-\e N$ due to the massive
matters. 
We will show in the large $N$ and
continuous limit the partition function of the 2D Yang-Mills
theory is regarded as a large $N$ limit of the matrix model in some
sense.

\section{2D Yang-Mills theory as instanton counting}
%\hspace*{-\parindent}{\it Decomposition into a product group}

We first consider the large $N$ limit 
of the partition function (\ref{profile partition})
with fixing the extra parameter $\e$ finitely; 
\begin{align}
 Z_G(N\to\infty, \e, \lambda A) 
= \sum_{k=1}^{\infty}\sum_{\kv \in Y_k}
\exp\biggl\{
&-\frac{2-2G}{8}\pint dx dy\  f''_\kv(x|\e)f''_\kv(y|\e)\gamma_\e(x-y) \nn\\
&-\frac{\lambda A}{8 \epsilon^2}\int dx\ 
f''_\kv(x|\e) x^2
\biggr\}. 
\label{large N}
\end{align} 
Notice that the cubic potential term coming from the quadratic Casimir
and contribution of the ``cut-off'' matter part
has been dropped in this large $N$ limit.

In addition, let us consider a decomposition of the gauge group $U(N)$ into
a product group $U(N_1)\times U(N_2)\times \cdots \times U(N_r)$,
where $N=\sum_{i=1}^{r}N_i$.
To do this, we need to define preliminarily 
``sub-partitions'' of the original partition 
$\left\{k_i\right\}_{i=1}^{N}$ as follows; 
\begin{align}
 \kv^{(1)} 
  &= \left\{ k_{1,1}\cdots k_{1,N_1}\right\}
  \equiv\left\{ k_1, \cdots, k_{N_1} \right\},  \nn\\
 \kv^{(2)} 
  &= \left\{ k_{2,1}\cdots k_{2,N_2}\right\}
  \equiv \left\{k_{N_1+1},\cdots,k_{N_1+N_2}\right\}, \\ 
           &\qquad\qquad\qquad\qquad \vdots  \nn \\
 \kv^{(r)} 
  &= \left\{ k_{r,1}\cdots k_{r,N_r}\right\}
  \equiv \left\{k_{N_1+\cdots +N_{r-1}},\cdots, k_{N}\right\}. \nn
\end{align}
Then we can prove the relation, 
\begin{align}
 \prod_{i\ne j}^{N}\frac{k_i-k_j+j-i}{j-i} 
 &= \prod_{l=1}^r \prod_{i\ne j}^{N_l}
    \frac{k_{l,i}-k_{l,j}+j-i}{j-i} \nn\\
     &\quad\times
    \prod_{l\ne n}^{r}\prod_{i=1}^{N_l}\prod_{j=1}^{N_n}
    \frac{M_l-M_n+k_{l,i}-k_{j,n}+j-i}{M_l-M_n+j-i} \nn \\
 &= \prod_{(l,i)\neq(n,j)}
    \frac{M_l-M_n+ k_{l,i}-k_{n,j}+j-i}{M_l-M_n+j-i}, 
 \label{1st step}
\end{align}
where 
\begin{align}
 M_1 &= N, \nn \\
 M_2 &= N-N_1, \\ 
     &\quad \vdots \nn\\
 M_r &= N-N_1-\cdots -N_{r-1}. \nn
\end{align}
By setting $a_l\equiv \e M_l$, 
we can rewrite the partition function 
(\ref{large N}) as 
\begin{align}
 Z_G = \sum_{k=1}^\infty e^{-\frac{k\lambda A}{2}}
 \sum_{\kv \in Y_{\kv}} 
 \left(
 \prod_{(l,i)\neq(n,j)}
 \frac{a_l-a_n+\e(k_{l,i}-k_{n,j}+j-i)}{a_l-a_n+\e(j-i)}
 \right)^{\frac{2-2G}{2}}. 
 \label{partition function2}
\end{align}
For $G=0$, this is equal to the non-perturbative part of 
the instanton partition function 
of 4D ${\cal N}=2$ supersymmetric gauge theory derived 
in Ref.\,\cite{Nekrasov:2003af}, 
except that the centers of $U(1)$, $a_l$'s, are freezed and
rigidly related to each $N_i$ in this construction.

\begin{figure}[ht]
\begin{center}
\includegraphics[scale=0.8]{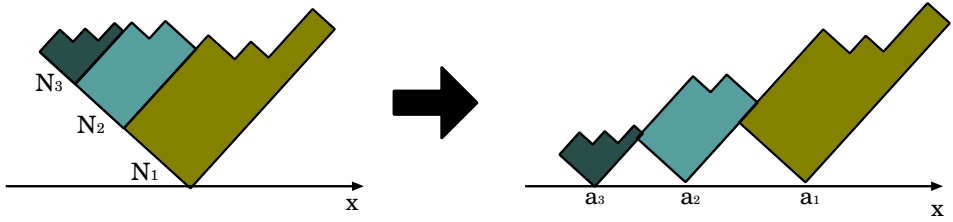}
\end{center}
\caption{The decomposition of the $U(N)$ Young diagram into the irreducible
representation of the product group $U(N_1)\times U(N_2)
\times U(N_3)$ as an example,
 where $N_1+N_2+N_3=N$. The Young diagram divides into
 three pieces and the origins of the profile function sit at $x=a_1,a_2,a_3$.}
\label{divided profile}
\end{figure}

This observation suggests that 2D Yang-Mills theory 
should reproduce the instanton counting of 
the 4D ${\cal N}=2$ supersymmetric gauge theory 
by considering the gauge symmetry breaking 
from $U(N)$ to the product group $\prod_{i=1}^{r}U(N_i)$. 
After the gauge symmetry breaking, 
we can freely move overall $U(1)$ charges for each $U(N_i)$ factor, 
that is, we can deal with $a_l$'s as free parameters. 
From the point of view of the Young diagram, 
it is done by sliding each sub-diagram down to the ``ground line'' 
and moving $x$-direction freely in the profile function, 
since irreducible representations of the product group is
embedded in a Young diagram of the original $U(N)$ with 
$N=\sum_{i=1}^{r}N_i$ as sub-diagrams. 
(See Fig.\ref{divided profile}.) 

In the language of the profile function it is realized 
by replacing $f_{\kv}(x|\e)$ in (\ref{profile partition}) 
to the ``colored partition'' \cite{Nekrasov:2003af},
\be
f_{\av;\vec{\kv}}(x|\e)
\equiv\sum_{l=1}^{r} f_{\kv_l}(x-a_l|\e), 
\label{deformation}
\ee 
with centers of the profiles $a_l$.
The positions of the origins of the profile function are regarded as the
 ``$U(1)$ charge'' of each $U(N_i)$ factor. Actually, if we consider the
reduction of the Young diagram from $U(N)$ to $SU(N)$, we need to tear off
the rectangle block with the width of the last row $k_N$, 
thus the $SU(N)$ Young
diagram only has up to $N-1$ rows. 
This operation makes a shift of the origin of the profile
function by $k_N$ and the rectangle block 
corresponds to the overall $U(1)$ charge. 
Therefore, the positions of the origins, $a_l$'s, now become free parameters 
since we can change the $U(1)$ charges of $U(N_l)$'s, 
although $a_l$'s in (\ref{1st step}) seem to have fixed values. 
Note that 
this modification of 2D Yang-Mills theory
would be explained by adding an adjoint scalar 
field to the 2D theory and integrating 
out off-diagonal massive components,  
which is naturally understood 
if we consider a brane configuration 
that realizes 2D Yang-Mills theory on $S^2$. 
(For detail, see the section 5.) 

After the modification (\ref{deformation}),
we find that the large $N$ limit of 
the partition function (\ref{large N}) becomes
\begin{equation}
 Z_G = Z_{\rm pert}^{\frac{2-2G}{2}}(\av,\e)
  \sum_{k=1}^{\infty} e^{-\frac{k\lambda A}{2}}
  \sum_{\kv \in Y_k} \mu_{\kv}^{2-2G}(\av,\e), 
  \label{instanton partition}
\end{equation}
where
\bea
Z_{\rm pert}(\av,\e) &=& \exp\left\{
\sum_{l \neq n}\gamma_\e(a_l-a_n)
\right\},\\
\mu_{\vec{\kv}}^{2}(\av,\e)
&=&\prod_{(l,i)\neq(n,j)}
\frac{a_l-a_n+\e(k_{l,i}-k_{n,j}+j-i)}{a_l-a_n+\e(j-i)}. 
\eea
Therefore, if we choose 2D Yang-Mills theory on a sphere ($G=0$),
the partition function exactly agrees with Nekrasov's partition function
by an identification of $q=e^{2\pi i \tau}=e^{-\frac{\lambda A}{2}}$.%
\footnote{The area $A$ can be complexified 
by turning on the $\theta$-angle ($U(1)$ flux) of $U(N)$ theory.}
In the following section, we discuss why this happens 
from the string theoretical point of view.

Let us evaluate the free energy of 2D Yang-Mills theory 
in the large $N$ limit.
To derive an expansion of the free energy from 
the partition function (\ref{instanton partition}), 
it is convenient to rewrite as
\be
Z_{G=0}(N\rightarrow \infty;\av,\e)
=Z_{\rm pert}(\av,\e)
\left(
1+\sum_{k=1}^{\infty}
q^k Z_k(\av,\e)
\right),
\label{large N partition}
\ee
where
\bea
Z_{\rm pert}&=&\exp\left\{
\sum_{l \neq n}\gamma_\e(a_l-a_n)
\right\},\\
Z_k(\av,\e) &=& Z_{\rm pert}^{-1}
\sum_{\vec{\kv}\in Y_k}
\exp\left\{-\frac{1}{4}\pint dxdy\ 
f_{\av;\kv}''(x|\e)f_{\av;\kv}''(y|\e)\gamma_\e(x-y)\right\},
\eea
and $q \equiv e^{-\frac{\lambda A}{2}}$.
Nekrasov claims that the prepotential of $\N{=}2$ 
4D supersymmetric Yang-Mills theory can be obtained by
a continuous limit of the free energy of the large $N$ (planar)
partition function (\ref{large N partition}),
\cite{Nekrasov:2003af, Nekrasov:2002qd}
\bea
{\cal F}_{0} &\equiv& -\lim_{\e\rightarrow 0}
 \e^2\log Z_{G=0}(N\rightarrow \infty;\av,\e)\\
&=&{\cal F}_0^{\rm pert} + {\cal F}_0^{\rm inst},
\eea
where ${\cal F}_0^{\rm pert}$ and ${\cal F}_0^{\rm inst}$ stands
for the perturbative and non-perturbative instanton contribution to
the prepotential, and obtained explicitly
from the expansion (\ref{large N partition})
\bea
{\cal F}_0^{\rm pert} &=&
 -\lim_{\e\rightarrow 0}\ \e^2 \log Z_{\rm pert},\\
{\cal F}_0^{\rm inst} &=&
 -\lim_{\e\rightarrow 0}\ \e^2 \log \left(1+\sum_k q^k Z_k\right).
\label{Finst}
\eea
Indeed, using the expansion of $\gamma_\e(x)$ in $\e$
 (\ref{expansion of gamma}),
 we find the
 perturbative part of the prepotential up to rescaling by a constant
\be
{\cal F}_{0}^{\rm pert}
=\sum_{l \neq n}\left[
\frac{1}{2}(a_l-a_n)^2\log\left(\frac{a_l-a_n}{\Lambda}\right)
-\frac{3}{4}(a_l-a_n)^2
\right].
\ee
This agrees with the perturbative part of 
the Seiberg-Witten prepotential 
of $\N{=}2$ $SU(r)$ gauge theory \cite{Seiberg:1994aj}.
For the non-perturbative part, we can re-expand 
as a formal power series in $q$
since the summation in the logarithmic of (\ref{Finst}) starts from 1.
As a result, we obtain the expansion, 
\be
{\cal F}_0^{\rm inst} = \sum_{k=1}^{\infty} q^k {\cal F}_{0,k}.
\ee

We have taken the planar limit of the partition function in the limit of
$\e\rightarrow 0$, but we can also expand asymptotically
the whole 2D Yang-Mills free energy in general as follows
\be
{\cal F}_{\rm 2D YM}=
\sum_{g=0}^\infty \sum_{k=0}^{\infty}
\e^{2g-2}q^k{\cal F}_{g,k},
\label{expansion}
\ee
where we define ${\cal F}_{0,0}\equiv {\cal F}_{\rm pert}$.
This is a novel expression due to introducing an additional parameter $\e$.
From the 4D field theoretical point of view, the expansion in $\e$ indicates
a higher genus correction in a graviphoton background.
An essential meaning in 2D Yang-Mills theory is not so clear, but
the existence of the additional parameter $\e$ helps us to take
a well-regularized double scaling continuous limit as we will discuss
in the next section.

%%%%%%%%%%%%%%%%%%%%%%%%%%%%%%%%%%%%%%%%%%%%%%%%%%%%%%%%%%%%
\section{Double scaling limit}

By rewriting the partition function of 2D Yang-Mills theory  
using profile functions as (\ref{profile partition}), 
we can take various limits of the theory 
since it depends not only on $N$ and $\lambda A$ 
but also on the cell size of the Young diagram $\e$.
In this section, we take a ``double scaling limit'', 
\begin{equation}
N\to\infty, \quad \e\to 0, \qquad {\rm with}\ \ \e N = m = {\rm fixed}.
\label{double scale limit}
\end{equation}
In this limit, we find that the profile function becomes 
a smooth function. 
In fact, by setting 
\begin{equation}
\e i \equiv t, \qquad \e k_i \equiv k(t), 
\end{equation}
and defining a function $\th(t)$ as 
\begin{equation}
\th(t) \equiv -k(t)+t, 
\end{equation}
we see that the smooth limit of the profile function is given by 
\begin{equation}
\lim_{\e\to 0} f_{\kv}(x|\e) \equiv f_{k}(x) = 
\begin{cases}
\max\left( |x|, \ |x|+2(\th^{-1}(-x)+x) \right) &\qquad
(x <0) \\
|x| + 2 \th^{-1}(-x) &\qquad 
(x \ge 0) \\
\end{cases}. 
\end{equation} 
The second derivative of the profile function 
in the smooth limit is 
\begin{equation}
f_{k}''(x) =
2\delta(x) + 2\int_0^m dt 
\Bigl[\delta'(x+t)-\delta'(x+\th(t))\Bigr] .
\label{continuous profile} 
\end{equation}

As mentioned in the previous section, 
the position of a profile function is related to 
the $U(1)$ charge.
In order to see this we note that 
the quadratic casimir of $U(N)$ representation can be
obtained by that of the $SU(N)$ with $U(1)$ charge 
$q=k+Nr$;
\bea
C_2(R,q)=C_2(R)+q^2/N , 
\eea
where $C_2(R)$ is given as (\ref{Casimir potential}).
This can be written in terms of the profile function as 
\bea
C_2(R,q)=\frac{N}{2 \epsilon^2}\int dx 
\left[
f''_k(x- \epsilon r)
\left(\frac{1}{2}x^2+\frac{1}{3\epsilon N}x^3\right)
\right]
-\frac{N}{2}r^2+\frac{r^3}{3} .
\eea
Under the limit with $\epsilon r =a$ is kept fixed, 
we have
\bea
C_2(R,q)=\frac{N}{2 \epsilon^2}\int dx 
\left[
f''_k(x- a)
\left(\frac{1}{2}x^2+\frac{1}{3m}x^3\right)
\right]
+const ,
\eea 
in which we can see the connection between the center of colored
profile and the $U(1)$ charge.

Combining and substituting these into the partition function 
(\ref{profile partition}) and shifting $\th(t)$ by a constant as 
$\th(t)\equiv h(t)+m/2$, 
we obtain 
\begin{equation}
 Z_G = \int {\cal D}h(t) \exp\left( -\frac{1}{\e^2} S_{\rm eff}[h(t)] \right), 
 \label{continuum partition}
\end{equation}
where 
\begin{equation}
 S_{\rm eff}[h(t)] = -\frac{2-2G}{2}\pint ds dt \log\left( h(s)-h(t) \right) 
  +\frac{\lambda A}{2m} \int dt \left[ h(t)-a \right]^2.
\end{equation}
If we set $G=0$, 
this is nothing but the effective action of 2D Yang-Mills theory 
discussed by Douglas and Kazakov in Ref.\,\cite{Douglas:1993ii}. 
Therefore we can conclude that the ``large $N$ limit'' taken in 
Ref.\,\cite{Douglas:1993ii} is the double scaling limit 
(\ref{double scale limit}).%
\footnote{In Ref.\,\cite{Douglas:1993ii}, $m$ is set to be $1$.} 

Let us briefly review the discussion in Ref.\,\cite{Douglas:1993ii}. 
Since we take the limit $\e\to0$, 
the saddle point of the effective action dominates in 
(\ref{continuum partition}).   
By taking variation for $h(t)$, we obtain (with setting $a=0$)
\begin{equation}
 \pint dy \frac{\rho(y)}{h(t)-y} 
 = \frac{\lambda A}{2m}h(t), 
   \label{saddle point}
\end{equation}
where $\rho(x)=\frac{d h^{-1}(x)}{dx}$ is the ``density'' 
of boxes of the Young diagram. 
The one-cut solution of (\ref{saddle point}) is 
Wigner's semi-circle,  
\begin{equation}
 \rho(x) = \frac{\lambda A}{2\pi m}\sqrt{R^2 - x^2},   
  \qquad
 \left(
  R^2 = \frac{4m^2}{\lambda A}
 \right) 
  \label{semi circle}
\end{equation}
where $R$ is determined by the condition, 
\begin{equation}
 \int \rho(x) dx = m. 
\end{equation}
Since $h(t)$ expresses a Young diagram, 
$\rho(x)$ must be equal to or less than $1$. 
Therefore, there is a critical area, 
\begin{equation}
 \lambda A_c = \pi^2,  
\end{equation} 
at which there appears a third order phase transition. 

\begin{figure}[ht]
\begin{center}
\includegraphics[scale=1.0]{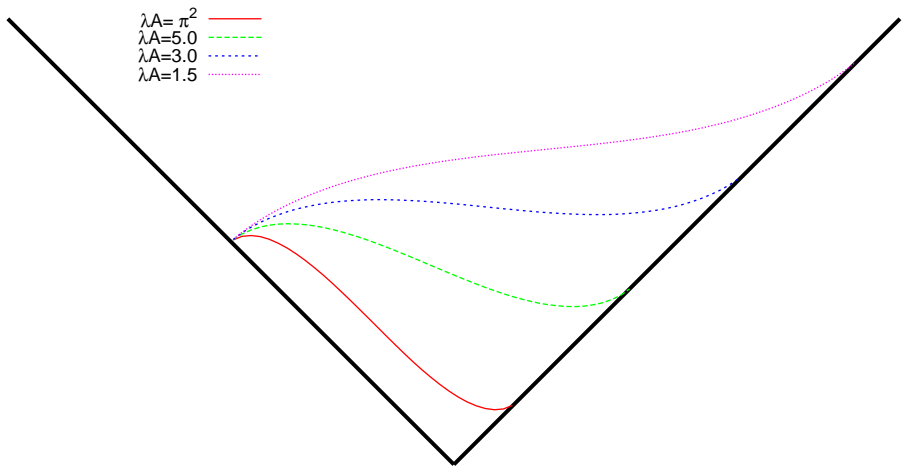}
\end{center}
\caption{The profile functions corresponding to the density 
satisfying Wigner's semi-circle law. 
As $\lambda A$ approaches to the critical area, 
$\lambda A_c=\pi^2$, the shape of the profile function 
becomes a rectangle. 
When the area becomes larger than the critical area, 
a line region would appear in the profile function.
Note that the critical area dose not depend on the value of $m$.  
}
\label{DK-profile}
\end{figure}

To see the shape of the Young diagram 
which is dominated in the double scale limit, 
it is convenient to determine the profile function corresponding to it.  
By integrating the density (\ref{semi circle}) for $x$, we obtain 
\begin{equation}
 h^{-1}(x) = 
  \begin{cases}
  0 &\qquad (x < -R) \\
  \displaystyle\frac{m}{\pi}\left[\arcsin\left(\frac{x}{R}\right)
  + \frac{x}{R}\sqrt{1-\left(\frac{x}{R}\right)^2} + \frac{\pi}{2} 
  \right]  &\qquad (-R \le x \le R) \\
  m  &\qquad (x > R)
  \end{cases}. 
  \label{profile in DK}
\end{equation} 
Substituting this into the continuous profile function 
(\ref{continuous profile}), 
we can easily exhibit the shape of Young diagrams 
dominating in the double scaling limit. 
In Fig.{\ref{DK-profile}}, we draw the shape of 
the profile function corresponding to the expression $\th(x)$ 
above.
Here profiles are obtained with choosing appropriate 
$U(1)$ charge or the center $a$ 
to slide the profile function towards the oblique direction 
in order that the profiles fall into place. 

In the figure, we see that a region where 
the derivative of the profile function 
is less than $-1$ appears when the area exceeds the critical area 
$\lambda A_c = \pi^2$. 
If we assume that this region is replaced by a line, $-x+const.$, 
we can understand this phase transition as a transition from 
the one-cut solution to a two-cut solution in the language of the
density $\rho(x)$  \cite{Douglas:1993ii}.

%%%%%%%%%%%%%%%%%
\section{Aspects from string theory}

We have seen in the previous sections 
that a connection between 2D Yang-Mills theory
on a sphere and the partition function of the 4D instanton
counting. This connection is so mysterious only from the field theoretical
point of view. However if we realize the 4D gauge theory
in string theory using branes, the connection becomes to be clear.

Let us now consider $N$ D5-branes wrapping 
on a 2-cycle ($\CP^1$) in an ALE space 
with resolved $A_1$ singularity in Type IIB superstring theory. 
This configuration preserves 8 supercharges on $\R^{1,3}$
worldvolume direction of D5-branes except for the 2-cycle and $\N{=}2$ $U(N)$
4D gauge theory appears.
And also, the internal theory on $\CP^1$ must be topologically twisted
and we expect that it is equivalent to a bosonic topological 
2D Yang-Mills theory of the BF type as discussed in \cite{Dijkgraaf:2003xk}
\be
S=\int_{\CP^1}\Phi F.
\ee

However, from the analysis of the double scaling limit in the section 4,
the limit of 2D Yang-Mills has the quadratic potential as
the $c=0$ matrix model below the Douglas-Kazakov phase transition point.
This means that the action might be deformed to
$\N{=}1$ by an induced quadratic potential
\be
S=\int_{\CP^1}\Phi F+\mu \Tr \Phi^2.
\ee
This theory reduces to an ordinary 2D Yang-Mills theory
after integrating out $\Phi$. Geometrically, the 2-cycle $\CP^1$ now turns
into a 2-cycle in the resolved conifold.

The gauge coupling of the 4D theory is proportional to an area
of the 2-cycle by a dimensional reduction, but the coupling could be
complexified
by adding the NS-NS B-field ($U(1)$ gauge field) through the 2-cycle 
which corresponds to the theta angle $\theta$.
Namely, the gauge coupling of 4D theory and 2D objects are related each other
by
\be
2\pi i \tau = -\frac{\lambda A}{2}+ i \theta.
\ee

In addition, the D5-branes
have extra two dimensional transverse directions, 
which is identified with the vev of the adjoint scalar 
in the vector multiplet of $\N{=}2$ theory
if we dropped the quadratic potential part in the large $N$ limit without
fixing $\e N$.
The situation of the product group we have considered in section 3
corresponds to $N_l$ $(l=1,\ldots,r)$ D5-branes are localized at the same place
of the transverse directions and the center of the D5-brane bunch is $a_l$.
We need to take each large $N_l$ limit  with fixing the center of positions
and require that the near horizon radius does not overlap each other
(each bunch should be sufficiently separated).

\begin{figure}
\begin{center}
\includegraphics[scale=0.7]{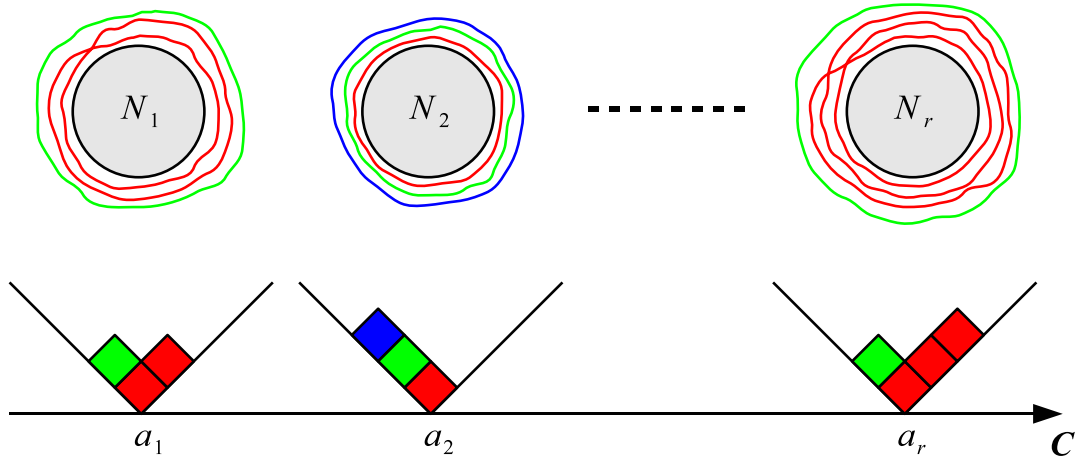}
\end{center}
\caption{A configuration of instantons (wrapping D-strings).
D-strings associated with the Young diagram are wrapping on $\CP^1$
where D5-branes are wrapping on the same cycle.}
\label{D-branes}
\end{figure}

The instanton correction comes from the (euclidean) D-string wrapping around
the blow-up 2-cycle. 
The $k$ instanton contribution corresponds to $k$ times
wrapping D-string. Using the string like description of the 2D Yang-Mills
partition function \cite{Gross:1993tu,Gross:1993hu,Gross:1993yt},
wrapping maps from the world-volume of D-string to the target $\CP^1$
are specified by the Young diagram (and related cycles). The Young diagram
of each gauge factor is located at $a_l$ 
\footnote{This position also holomorphically extends to a complex
in the large $N$ limit.}
and it describes how to wrap D-strings
on each localized $\CP^1$ of the $N_l$ D5-branes. (See Fig.\ref{D-branes}.)
The partition function (\ref{instanton partition}) counts all possible configurations
of the wrapping D-strings.

According to the gauge/geometry (open/closed string) correspondence
\cite{Maldacena:1998re,
Gopakumar:1998vy,Gopakumar:1998ki,
Dijkgraaf:2002fc,Dijkgraaf:2002vw,Dijkgraaf:2002dh},
the large number limit of D-branes
causes the geometry transition and D-brane charges turn to a
non-trivial background flux.
The flux carries all information of 4D $\N{=}2$ theory, that is,
the information of the Seiberg-Witten geometry is encoded in 
the R-R and NS-NS B-filed flux configuration. This is regarded as
a T-dual (along $x^6$) picture of the Hanany-Witten brane configuration
\cite{Hanany:1997ie, Witten:1997sc} in the large $N$ limit and also may
relate to topological (closed) string theory on various geometry.

\section{Conclusions and discussions}

In this article, we discussed 2D Yang-Mills theory 
using the technique of the random partition. 
We found that the partition function of the 2D Yang-Mills theory 
can be written in terms of a ``profile function'' $f_\kv(x|\e)$ 
which expresses the Young diagram corresponding to the partition $\kv$.
We examined two kinds of limits of the theory; 
the large $N$ limit and the double scaling limit. 
We showed that the large $N$ limit of the partition function 
reproduces the partition function of 
the instanton counting of the 4D $\N=2$ 
supersymmetric gauge theory discovered by Nekrasov. 
On the other hand, we found that the double scaling limit of 
the partition function realizes naturally the discussion in 
Ref.\,\cite{Douglas:1993ii} by Douglas and Kazakov. 
We also gave an interpretation of the instanton counting 
from the view point of brane configurations in the superstring theory. 

We conclude this article by making some comments. 
In the section 3, we took the large $N$ limit of 
the 2D Yang-Mills theory on $S^2$ reproduces the instanton counting 
of a 4D $\N=2$ supersymmetric gauge theory. 
From this fact, it seems natural to expect that the double scaling 
limit of the 2D Yang-Mills theory discussed in the section 4 
describes some non-perturbative aspects of a 4D theory. 
In fact, if the area is smaller than the critical area, 
the double scaling limit of the same theory is closely related to 
a $c=0$ matrix model. 
It suggests that the double scaling limit of 2D Yang-Mills theory 
might describe non-perturbative aspects of a 4D 
$\N=1$ supersymmetric gauge theory \cite{Dijkgraaf:2002dh}. 
The deference between the partition function of 
the large $N$ limit and the double scaling limit 
is the cubic potential term $\int dx f_{\kv}''(x|\e) x^3$. 
If the cubic potential turns on, the ALE space where D5-branes live 
might be deformed to a Calabi-Yau manifold, 
which reduces the supersymmetry 
on the 4D space-time from $\N=2$ to $\N=1$.
Moreover, if we admit this assumption, 
the phase transition discussed in Ref.\,\cite{Douglas:1993ii} 
could be understood from the view point of the brane picture. 
In this picture, the area of $S^2$ where 2D Yang-Mills theory lives 
is that of a resolved 2-cycle in a CY manifold. 
When the area is small, the low energy theory on the D5-branes 
would be a 4D $\N=1$ supersymmetric gauge theory, 
which is consistent with the fact that the effective action 
of the 2D Yang-Mills theory is that of a $c=0$ matrix model 
in this parameter region. 
However, if the area becomes large enough, 
the effective theory on the D5-branes would not be a 4D theory 
but a 6D theory. 
The critical area might be identified with the area 
at which we cannot ignore the size of the $S^2$. 

As for the phase transition, Gross and Witten have also discussed 
that a third order phase transition appears in 
the large $N$ limit of the one-plaquette model \cite{Gross:1980he}. 
Recently, the authors in Ref.\,\cite{deHaro:2004id} have shown, 
among other things, that Willson's one plaquette model
can be formulated via the method of one-dimensional discrete random
walk model which is also related to growing Young diagram, 
in which the authors argue that the Gross-Witten 
third order phase transition occurs
when the growing Young diagram reaches its ceiling. 
It seems of quite interest to explore 
the relationship between the phase transition in 
the large $N$ limit of the one-plaquette model discussed 
by Gross and Witten \cite{Gross:1980he} 
and the Douglas-Kazakov phase transition discussed 
in the section $4$.

\section*{Acknowledgements}
The authors would like to thank H.~Kawai, T.~Tada, M.~Hayakawa, 
T.~Kuroki, Y.~Shibusa and T.~Sakai
for useful discussions and valuable comments. 
TM would like to thank H.~Kanno and H.~Itoyama for 
valuable comments. 
KO also would like to thank N.~Dorey, T.~Hollowood and H.~Fuji
for useful conversations.
This work is supported by Special Postdoctoral Researchers
Program at RIKEN.

\bibliographystyle{JHEP}
\bibliography{refs}

\end{document}